\documentclass[12pt,a4paper]{article}
\usepackage{amsmath}
\usepackage{amssymb}
\usepackage{bm}
\usepackage{graphicx}
\usepackage{rotating}
\usepackage{pdfsync}
\usepackage{multirow}
\usepackage{natbib}
\usepackage{har2nat}
\usepackage{pdflscape}
\usepackage{setspace}
\usepackage{xcolor} %for color
\usepackage[colorlinks,bookmarksopen,bookmarksnumbered,citecolor=red,urlcolor=red]{hyperref}

\title{Construction and assessment of prediction rules for binary outcome in the presence of missing predictor data using multiple imputation: theoretical perspective and data-based evaluation.}
\author{Mertens, B. J. A.$^1$, Banzato, E.$^2$, de Wreede, L.C.$^1$}

\begin{document}
%\date
\maketitle\vspace{-0.5cm} \noindent $^1$Medical Statistics,  Department of Biomedical Data Sciences, Leiden University Medical Center, PO Box 9600, 2300 RC
Leiden,  Netherlands\\
$^2$Department of Statistical Sciences, University of Padova, Padova, Italy.
\\
\mbox{}

%\runninghead{Mertens, Banzato and de Wreede}
%\title{Accounting for uncertainty in the predictive calibration of prognostic models constructed using multiple imputations with cross-validatory assessment.
%%On the combination of cross-validation and multiple imputation in clinical survival prediction modeling.
%}
%\author{Mertens, B. J. A.\affilnum{1}, Banzato, E.\affilnum{2}, de Wreede, L.C.\affilnum{1}}
%\affiliation{\affilnum{1}Medical Statistics,  Department of Biomedical Data Sciences, Leiden University Medical Center, PO Box 9600, 2300 RC
%Leiden,  Netherlands\\
%\affilnum{2}Department of Statistical Sciences, University of Padova, Padova, Italy.}
%\corrauth{Mertens, B. J. A.,
%Medical Statistics,  Department of Biomedical Data Sciences, Leiden University Medical Center, PO Box 9600, 2300 RC
%Leiden,  Netherlands}
%\email{b.mertens@lumc.nl}

\begin{abstract}
We investigate the problem of calibration and assessment of predictive rules in prognostic designs when missing values are present in the predictors.
Our paper has two key objectives which are entwined. The first is to investigate how the calibration of the prediction rule can be combined with the use of multiple imputation to account for missing predictor observations.
The second objective is to propose such methods that can be implemented with current multiple imputation software, while allowing for unbiased predictive assessment through validation on new observations for which outcome is not yet available.

To inform the definition of methodology, we commence with a review of the theoretical background of multiple imputation as a model estimation approach as opposed to a purely algorithmic description.
We specifically contrast application of multiple imputation for parameter (effect) estimation with predictive calibration.
Based on this review,  two approaches are formulated, of which the second utilizes  application of the classical Rubin's rules for parameter estimation,  while the first approach averages probabilities from models fitted on single imputations  to directly approximate the predictive density for future observations.
We present implementations using current software which allow for validatory or cross-validatory estimation of performance measures,  as well as imputation of missing data in predictors on the future data where outcome is by definition as yet unobserved.

To simplify the discussion we restrict discussion to binary outcome and logistic regression throughout, though the principles discussed are generally applicable.
We present two data sets as examples from our regular consultative practice.
Method performance is verified through application on the real data.
We specifically investigate accuracy (Brier score) and variance of predicted probabilities.
Results show little difference between methods for accuracy but substantial reductions in variation of calibrated probabilities when using the first approach.

\end{abstract}

\maketitle

\section{Introduction}
There has been much recent interest in the  medical statistical, epidemiological and even biomedical literature on calibration and validation of prediction rules in the prognostic context, when multiple imputations are used to account for missing observations in the predictors.
This renewed interest has been supported by 1) the emergence of easy-to-use packages for the generation of imputations within statistical software such as {\tt R} or {\tt Stata},  as well as 2) the now ready availability of fast and cheap computing
even with desktop configurations.
This has unleashed
creativity to propose and investigate various novel combinations of predictive calibration with validation approaches and imputation.

A feature of this literature is that it is predominantly algorithmic and somewhat ad hoc in nature.
Wood {\it et al.} (2015) for example  focus on performance assessment and formulates various strategies.
Wahl {\it et al.} (2016) investigate the problem of combining predictive calibration with validation and imputation but with a particular focus on bootstrapping.
Generation of post hoc summaries,  among which performance estimates, after multiple imputation, is discussed by Marshall {\it et al.} (2009).  These authors also report on a literature review in  recent biomedical literature on  use of multiple imputation in prognostic studies.
Vergouwe {\it et al.} (2010) discuss  case studies on practical development of prognostic models with imputation,   also addressesing model selection.
Different strategies for predictive calibration with imputation are discussed by Miles (2015).

In line with the predominantly algorithmic nature of these presentations, there is little attempt to tie  proposed algorithmic development to an established theoretical framework (see \cite{Carpenter2013}, \cite{Carlin2015} for a thorough review of the theoretical background of multiple imputation e.g.,  among many other such sources).
Novel methods are developed as adaptations of or combinations with the multiple imputation algorithm.
Indeed,  multiple imputation itself tends to be presented as an algorithmic device, although it has a clear methodological foundation as an approximation of the joint density of effect estimates  near the mode.
Multiple imputation is {\bf \it model} estimation.
By re-establishing focus on multiple imputation as a model approximation and thus estimation approach, it may become more easy to identify suitable approaches for method validation by formulating validation as model assessment and similarly for the definition of the predictive approach itself.

There is consensus within the literature on the fundamental challenge posed by multiple imputation in prognostic calibration,  which is that while imputation must take into account observed outcomes,  unbiased validation by definition requires outcomes to be omitted when generating predicted values (\cite{Wood2014}, page 615; \cite{Wahl2016}, page 2).
A less recognized issue is that multiple imputation in the predictive context requires calibration of a distinct predictive  density than is currently allowed for with existing software which is focused on effect estimation instead.

To elucidate these issues,  our paper commences with a review and clarification of the theoretical foundation of multiple imputation with special emphasis on the distinction between prediction and effect estimation in the imputation context (section 2).
Based on this discussion,  we propose two basic approaches for the calibration of prognostic rules with multiple imputations which can be implemented with existing imputation software and allows for validation using a set-aside test set excluding outcome data (section 3). The second of these is based on classical Rubin's rule estimation,  while the first utilizes an approximation to the predictive density of future outcome.  This discussion may be viewed as a formalisation of the methods suggested by Miles (2015).
In contrast to the above discussed existing literature which predominantly focuses on simulation (Vergouwe's paper being a notable exception),  we subsequently present a data-based application using two real datasets from our own consultative experience  which have motivated our interest in this research. To compare methods, we study {\it \bf data-based} summary statistics,  specifically predictive accuracy and variance  based on application of methods to the data (sections 4 and 5).
We finish with a review of main results, key conclusions and formulate recommendations.

\section{Theoretical perspective}
\subsection{Parameter estimation and Rubin's rules}
To formalize our discussion, we assume  a substantive  prediction model $f(Y\mid \bm{X},\bm{\beta})$, which describes the variation in a univariate outcome $Y$ of interest,  conditional on a vector of predictors $\bm{X}$ and depending on an unknown vector of regression parameters $\bm{\beta}$.  The latter will need to be estimated using a sample from the population, prior to subsequent use of the model.
We only consider scenarios with missing data in the predictors in this paper,  such that $\bm{X}=(\bm{X}_m,\bm{X}_o)$, which separates into missing $\bm{X}_m$ and observed $\bm{X}_o$ components and with $Y$ fully observed.  If our primary interest were to reside in the regression parameter vector $\bm{\beta}$, then we would seek to estimate the conditional (so called a-posterior) density
\begin{equation} \label{eq1}
\begin{split}
p(\bm{\beta}\mid \bm{X}_o, Y)&=\int p(\bm{\beta}, \bm{X}_m \mid \bm{X}_o, Y) d\bm{X}_m\\
&=\int p(\bm{\beta} \mid \bm{X}_m, \bm{X}_o, Y) p(\bm{X}_m \mid \bm{X}_o, Y) d\bm{X}_m
\end{split}\end{equation}
which is obtained as
the marginalized joint a-posterior density on the two unknown components $\bm{\beta}$ and $\bm{X}_m$, marginalized across the nuisance unobserved covariate values in $\bm{X}_m$. The last equality reveals  this may also be thought of as the probability density for the target parameter of interest $\bm{\beta}$, conditional on the unknown quantities $\bm{X}_m$, averaged across the uncertainty in $\bm{X}_m$ (but always conditional on the actually observed data). This latter equality reveals the workings of classical multiple imputation, as it generates imputed data from the conditional density $p(\bm{X}_m \mid \bm{X}_o, Y)$, for each of which simulations may be generated from the corresponding densities $p(\bm{\beta} \mid \bm{X}_m, \bm{X}_o, Y)$, to approximate the moments of $p(\bm{\beta}\mid \bm{X}_o, Y)$ in a sampling-based manner.
The Rubin's rules-based approach represents a practical compromise to achieve this averaging,
by first sampling imputations
$\widehat{\bm{X}}_{m,k}$ drawn from $p(\bm{X}_m \mid \bm{X}_o, Y)$ and with $k=1,...,K$ for a total number of $K$ imputations.
Subsequently,  we
estimate the modes $\widehat{\bm{\beta}}_k$
of the conditional densities $p(\bm{\beta} \mid \widehat{\bm{X}}_{m,k}, \bm{X}_o, Y)$ evaluated at the completed datasets $(\widehat{\bm{X}}_{m,k}, \bm{X}_o, Y)$ and for all $k$.
Large-sample results from classical frequentist  theory  are then used to approximate the conditional density $p(\bm{\beta}\mid \bm{X}_o, Y)$ at the mode and these results give rise to the so-called Rubin's rule estimate of the expectation as
\begin{equation}\label{eq2}
\widehat{\bm{\beta}}_{MI}=\frac{1}{K} \sum_{k=1}^{K} \widehat{\bm{\beta}}_k.
\end{equation}
Readers can consult \cite{Carpenter2013}, pages 46-48, \cite{Carlin2015} or \cite{Gelman2004} pages 519-523, among many other sources  for further results, details and different perspectives on the approach.

\subsection{Prediction, the predictive density and imputation}
In the predictive scenario, the averaging described in equation~\ref{eq1}  no longer suffices and should be expanded to  average across the regression coefficients, in order to account for {\it both} the missing values $\bm{X}_m$,  and the uncertainty in $\bm{\beta}$.
Let $\widetilde{Y}$ be a future univariate outcome, which we want to predict from covariates $\widetilde{\bm{X}}$.
As before, we have available a previous sample of data from the sample population, with outcomes $Y$ and covariates $\bm{X}$,  which we will refer to as the calibration data.
To simplify the discussion and notations, we will in the first instance assume that there are no further missing values in the predictor data $\widetilde{\bm{X}}$,  such that we can write
$p(\widetilde{Y}\mid\bm{X}_o, Y)$ for the predictive density of future outcomes, which denotes the conditional dependence on the past observed calibration data $\bm{X}_o, Y$,  while ignoring the obvious dependence on $\widetilde{\bm{X}}$ for the time being.

In analogy to the previous section,  the predictive density must be calibrated
 for future outcomes $\widetilde{Y}$ as
\begin{equation}\label{eq3}
\begin{split}
p(\widetilde{Y}\mid\bm{X}_o, Y)&=\int f(\widetilde{Y},\bm{\beta},\bm{X}_m\mid\bm{X}_o, Y)d\bm{\beta}d\bm{X}_m\\
&=\int f(\widetilde{Y}\mid\bm{\beta},\bm{X}_m,\bm{X}_o, Y) p(\bm{\beta},\bm{X}_m\mid \bm{X}_o, Y)d\bm{\beta}d\bm{X}_m.
\end{split}
\end{equation}
The last line shows that the integration can now be achieved by averaging across both imputations $\widehat{\bm{X}}_{m,k}$ and simulations $\widehat{\bm{\beta}}_k$  from the density $p(\bm{\beta},\bm{X}_m\mid \bm{X}_o, Y)$, while conditioning on the observed calibration data $\bm{X}_o, Y$.
In analogy to equation~\ref{eq1}, this  implies we may calculate the expectations
$\widehat{P}_k=E(\widetilde{Y}\mid\widehat{\bm{\beta}_k},\widehat{\bm{X}}_{m,k},\bm{X}_o, Y)$ for each pair of imputed values
$\widehat{\bm{\beta}_k},\widehat{\bm{X}}_{m,k}$, from the conditional density $p(\bm{\beta},\bm{X}_m\mid \bm{X}_o, Y)$. The set of predictions $\widehat{P}_k$
%(\widetilde{Y}\mid\bm{X}_o, Y)
, $k=1,...,K$, may then be summarized using the mean in analogy to Rubin's rules, medians or some other suitable summary measure to get the final prediction estimate $\widehat{P}$. For example,  using Rubin's rules to summarize the set of predictions $\widehat{P}_k$, $k=1,...,K$, will estimate $E(\widetilde{Y}\mid\bm{X}_o, Y)$ as
\begin{equation}
\label{eq4}
\widehat{P}_{MI}=\frac{1}{K} \sum_{k=1}^{K} \widehat{P}_k.
\end{equation}
%We note that a na\"ive application of Rubin's rules would instead most likely propose to use the estimates
%\begin{equation}
%\label{eq5}
%\widehat{p}_{RB}=\int f(\widetilde{Y}\mid \widehat{\bm{\beta}}_{MI},\widetilde{\bm{X}}_m) p(\widetilde{\bm{X}}_m\mid \widetilde{\bm{X}}_o,\bm{X}_o, Y)d\bm{X}_m d\widetilde{\bm{X}}_m
%\end{equation}

For full generality,  the future outcomes may themselves also have missing values in the predictors,  such that $\widetilde{\bm{X}}=(\widetilde{\bm{X}}_o,\widetilde{\bm{X}}_m)$, and  remembering  that the actual missing observations may not occur in the same covariates containing missing values in the calibration data.
%The above notation has suppressed the dependence of the predicted probabilities $\widehat{p}_k$ and $\widehat{p}$ on the covariates values $\widetilde{\bm{X}}$ in the new data, where predictions should be obtained, for ease of notation.
In the presence of missing values, we will have in full generality that
\begin{equation}
\widehat{P}=E(\widetilde{Y}\mid \widetilde{\bm{X}}_o,\bm{X}_o, Y),
\end{equation}
and similarly for the $\widehat{P}_k$, which implies that the above equation~\ref{eq3} should also be expanded in the obvious manner to include averaging across $\widetilde{\bm{X}}_m$.
%In the general scenario where $\widetilde{\bm{X}}$ itself  contains missing values and thus the set $\widetilde{\bm{X}}_m$ is non-empty, this implies an additional non-trivial complication as the uncertainty due to these missing values
%should also be accounted for when predicting the outcome in the new data.
Furthermore,  there is an additional non-trivial complication if we wish to use the predicted outcomes $\widetilde{Y}$ to assess the predictive capacities of any approach in the presence of missing data $\widetilde{\bm{X}}_m$, as it is essential that any imputation model used for the unobserved components of $\widetilde{\bm{X}}$ does not make use of the associated outcomes $\widetilde{Y}$.
This would apply particularly for cross-validation, but seems to generate a conflict between multiple imputation and cross-validation, as outcomes are needed in any implementation of multiple imputation to preserve the correlation structure with the outcomes to be predicted.

%Several options may exist to resolve this problem,  on of which is to use separate imputation models for the calibration set $(Y, \bm{X}$ and the validation samples in $\widetilde{Y}, \widetilde{\bm{X}}$,  but this will typically be cumbersome as the validation sample sizes may typically be (much) smaller. The ideal solution would in principle seem to be to calibrate an imputation model on the calibration data only,  and of which the substantive prediction model for the primary outcome is  a component. This imputation model may then be applied to the set-aside validation data $\widetilde{Y}, \widetilde{\bm{X}}$,  and for both the prediction of the outcome and imputation of any unobserved covariate values,  but this requires existing multiple imputation software to have sufficient flexibility to allow full predictive model calibration and validation for the targeted outcome,  which does not presently seem to be the case.
%

\section{Methodological implementation using existing imputation software}
In principle,  implementation of the above approach is automatic and completely standard within the (fully) Bayesian approach. It has been amply described in the literature (\cite{Gelman2004}, \cite{Brooks}).
Summarizing for simplicity, it consists of calibrating the conditional densities of any predictor  variable,  conditional on all other predictor variables and the outcome. In addition and crucially,  we also need to calibrate the conditional density of the outcome conditional on all predictors,  which is the primary model component of interest in the predictive context. Missing values are treated as unknown parameters within this approach as discussed above and their estimation as well as that of any outcome, proceeds in an iterative fashion  starting from suitable starting values until convergence,  as in regular MCMC-based estimation, sequentially simulating values from the appropriate conditional densities. Optimization of this iterative sequence of equations constitutes calibration of the joint model on outcome and missing values from the primary (training) data. Once convergence is achieved,  the resulting system of equations may be applied to the set of predictor values of any new observation (for which the outcome has not yet been observed) and the simulated outcome measures may be suitably summarized to generate the predicted value.  The latter is essentially the approach taken in recent contributions by \cite{Erler2015} and \cite{Erler2017} for example, which is also a good recent illustration of the methodology.
This approach is likely  optimal from the predictive point of view. Nevertheless, it may still suffer from practical drawbacks.

\begin{enumerate}
\item The approach is intrinsically of much higher complexity than is customary in current traditional clinical application.  Some users may  have philosophical objections to the use of the fully Bayesian approach.

\item The method is difficult to implement and requires a high level of technical expertise and knowledge of Bayesian computing which will usually be lacking.

\item The Bayesian approach may be difficult to validate,  particularly in situations with small to medium sample sizes when a separate set-aside test set cannot be made available.  This applies particulary when cross-validation must be used.
\end{enumerate}

The last is probably the most serious, besides the need to abandon the traditional Rubin multiple-imputation compromise framework and associated software with which many researchers will be familiar. In the remainder of this paper we restrict to cross-validation and formulate an approach to approximate the predictive calibration described in section 2.2 as closely as possible using existing MI software,  while also allowing for cross-validation.

 To achieve this, we first describe a general approach to  validation which allows outcome data $\widetilde{Y}$ to be set-aside for subsequent validation of prediction rules,
 while also allowing for the imputation of any missing data $\widetilde{\bm{X}}_m$ and ${\bm{X}}_m$ in the corresponding validation and calibration predictor sets respectively (section 3.1). We then propose an algorithm which directly estimates the
outcomes by pooling predictions and contrast this with an alternative approach based on direct applications of Rubin's rule (section 3.2) for the estimation of model parameters.
 Although our discussion focuses  on cross-validation,  it could be adapted in an obvious manner for a single set-aside validation set.

\subsection{Combining cross-validation and multiple imputation}
A simple approach to set-aside outcome data and generate (multiple) imputations, while preventing the problems described end of section 1 and section 2.3, is to remove the complete set of outcomes $\widetilde{Y}$ from each left-out fold which is defined within the cross-validation.
Imputation models may then be fit on the remainder of the observed data $(\widetilde{\bm{X}}_o,\bm{X}_o, Y)$ and imputations can be generated from these models, including for any unobserved data $\widetilde{\bm{X}}_m$ in the left-out fold predictor set. In other words,  the outcomes are artificially set to `missing' within the set-aside validation fold. After imputation of the missing observations, a suitable prediction model can be fit on the imputed calibration data $(\widehat{\bm{X}}_{m},\bm{X}_o, Y)$. We then apply this model to predict the outcomes from the imputed validation predictor data $(\widehat{\widetilde{\bm{X}}}_m,\widetilde{\bm{X}}_o)$.
The outcomes $\widetilde{Y}$ are then returned to the left-out fold,  after which the entire procedure can be repeated for the next fold within the entire cross-validatory sequence.  Note that the imputed values for $\widetilde{Y}$ are simply discarded.

\subsection{Combining predictive calibration with multiple imputation}
With the above implementation of multiple imputation and validation,  there are two basic approaches to calibrate prediction rules with multiple imputations, while allowing for cross-validation assessment of predictions for the set-aside outcome data with existing MI software.

The first is to define the folds on the complete dataset, after which a {\it single imputation} and corresponding predictions for the set-aside outcomes are generated for each fold as described above.  This procedure generates a complete set of predictions across the entire dataset based on application of single imputation,  after which we may re-define the fold-definition and repeat the procedure. In this manner,  we generate a large set of predictions  $\widehat{\widetilde{Y}}_{ik}$,  across al observations $i=1,..,n$ and for $k=1,..,K$ for $K$ repetitions of the approach.  Prediction and {\it multiple} imputation are thus entwined in this approach and the final prediction can be derived by taking means or medians or other suitable summary across the $K$ predictions within each individual.

The second approach uses only a single fold definition which is kept fixed across multiple imputations. For each left-out fold in turn,  $K$ (multiple) imputations are then generated on the corresponding calibration and validation predictor data $(\widetilde{\bm{X}}_o,\bm{X}_o, Y)$,  after which Rubin's rule is applied to obtain estimates of the model parameters in a single consensus model.   The latter single model can then be applied to generate - in principle - predictions on the $K$ imputed predictor sets  $(\widehat{\widetilde{\bm{X}}}_m,\widetilde{\bm{X}}_o)$,  such that we have in full generality again $K$ predictions for each individual.   The latter will of course all coincide for complete records.

A fundamental difference between the first and second approach is that we use $K$ distinct models for the prediction of a single observation in the first,  while there is only a single (Rubin's rule combined) model used in the second method. The other difference is the extra variation in fold definitions in the first approach. Alternatively,  approach 1 can be seen as a compromise approximation to the calibration of the predictive density as described section 2.2 and which can be implemented using standard software. Approach 2 on the other hand uses the model
\[
f(\widetilde{Y}\mid \widehat{\bm{\beta}}_{MI},\bm{X})
\]
which is obtained by using the pooled (Rubin's rule) model parameters as plug-in point estimators in the assumed substantive population model.
Tables~\ref{table2A} and ~\ref{table2B} display the structure of both approaches in the case of logistic regression with multiple imputation and cross-validation for the analysis of binary outcome.  In addition to these two approaches,  we also investigated a third,  which is a variant of approach 2.  It consists of also averaging the imputations within the predictors (in addition to averaging the generated regression coefficients) within each individual, and then apply the pooled regression coefficient to the predictor data with missing values replaced by the averaged imputed values (\cite{Marshall}).

%\subsection{Na\"ive approaches to the combination of cross-validation and multiple imputation}
%In addition and analogy to the above  approaches, we also define so-called na\"ive implementations.  These consist of
%%simply calculating the Nelson-Aalen cumulative hazard estimate on the entire dataset first and then
% directly computing a set of $K$ multiple imputations on the complete data, ignoring any subsequent cross-validation. Once all instances of the $M$ imputed datasets have been obtained, we  define $L$ folds for each imputed dataset separately,  after which the above described process of calculating predictions for the set-aside data in the left-out folds is repeated in analogy to approach 1.  This process is repeated separately and with  a different fold definition for each of the imputed datasets,  after which we again average predictions within each individual across the imputations. This provides the analogue to the first approach.   The alternative is to again use a fixed fold definition across all $M$ imputed datasets and repeat the procedures for approach 2,  after which Rubin's rule is applied as described before to derive the consensus model and apply it to the left-out imputed data and take averages or similar.

\begin{sidewaystable}[h]
\normalsize
\begin{center}
\begin{tabular}{p{0.45\textwidth}}
\hline
\hline
Approach 1 \\

Define $K$ and repeat the following steps $K$ times \\
	
	\begin{itemize}
		
		\item[] Define $L$ folds for the CV procedure.
Select each fold in turn as the validation data,  keeping all other data as calibration set. For each such fold carry out the following computation.
		
		\begin{enumerate}
			
			\item Remove the outcome from the validation data.

\item Combine this outcome-deleted validation data with the corresponding calibration data. Generate a single imputation on this combined set.

			\item Fit a logistic regression model in the imputations-augmented calibration portion of this combined set.
		
			\item Derive predictions for subjects $i$ in the corresponding imputed validation set from this model, using the equation
			\[
			\widehat{P}_{i,k} = {exp\{\widehat{\bm{x}}_{i,k}\widehat{\bm{\beta}}_{k}\}}/(1+{exp\{\widehat{\bm{x}}_{i,k}\widehat{\bm{\beta}}_{k}\}}).
			\]
			
		\end{enumerate}
		
	\end{itemize}

\\
Compute the final prediction for each $i^{th}$ individual as the average  $\overline{P}_i$ across all $K$ predictions $\widehat{P}_{i,k}$.

\\

\hline
\hline
\end{tabular}
\caption{\normalsize\label{table2A} Algorithmic description of approach 1 for combination of multiple imputation with cross-validation using the logistic regression modelling for binary outcome. $K$ represents the number of imputations and $L$ denotes the number of folds in the cross-validation. Note that the folds get re-defined for each new imputation in this approach.}

\end{center}
\end{sidewaystable}

\begin{sidewaystable}[h]
\normalsize
\begin{center}
\begin{tabular}{p{0.45\textwidth}}
\hline
\hline
 Approach 2 \\

 Define $L$ folds for the CV procedure and select each $l^{th}$ fold in turn as the validation set, keeping all other data as calibration set. For each such fold carry out the following computation. \\

	\begin{enumerate}

			\item Remove the outcome from the validation data.
			
\item Combine this outcome-deleted validation data with the corresponding calibration data.

		\item Run $K$ imputations on this combined set.
		
		\item Fit separate logistic regression models on the training portion of each of these $K$ imputed datasets.
		
		\item Compute the average $\bar{\bm{\beta}}$ of the $K$ regression coefficients vectors from these models.

		\item Derive predictions for subjects $i$ in the imputed validation sets from the combined model, using the equation
		\[
		\widehat{P}_{i,k} = {exp\{\widehat{\bm{x}}_{i,k}\bar{\bm{\beta}}\}}/(1+{exp\{\widehat{\bm{x}}_{i,k}\bar{\bm{\beta}}\}})
		\]
		
	\end{enumerate}
	
\\

Compute the final prediction for each $i^{th}$ individual as the average  $\overline{P}_i$ across all $K$ predictions $\widehat{P}_{i,k}$.

\\

\hline
\hline
\end{tabular}
\caption{\normalsize\label{table2B} Algorithmic description of approach  2 for combination of multiple imputation with cross-validation using the logistic regression modelling for binary outcome. $K$ represents the number of imputations and $L$ denotes the number of folds in the cross-validation. Note that the folds get defined first in this approach and are then held fixed,  with the imputations run within folds.}

\end{center}
\end{sidewaystable}

\section{Data}
We consider two datasets from our personal statistical consultation experience to illustrate and assess the proposed methodologies. Both examples investigate variation in all-cause mortality. The first of these (CRT data), studies a population subject to increased cardiovascular risk which underwent cardiac resynchronisation therapy (CRT).
It may reasonably be assumed to represent a missing completely at random example,  as missing data is caused by failure of equipment.
This does not apply for the second dataset (CLL data), which studies  chronic lymphocytic leukemia (CLL) patients who had a hematopoietic stem cell transplant.
As the details of these data have been described elsewhere  (\cite{Hoke2017}, \cite{Schetelig2017a}, \cite{Schetelig2017b}), we only review the essential characteristics of the data and refer readers to the above papers for details.

The CRT data consists of a sample of 1053 patients, of whom 524 cases (50\%) had missing observations. These missing observations are furthermore almost completely concentrated in a single predictor variable (Lvdias), with negligible numbers of missing values in a restricted set of other predictors. Missing observations for Lvdias were due to failure of the measuring device,  which give some credence to the missing completely at random assumption.   The CLL data contains 694 records  of which 241 contained missing values (35\%) mainly scattered across 3 predictor variables. These are performance status (9\% missing), remission status (6\% missing) and cytogenic abnormalities (25\% missing).  For both data, the predictor set was pre-specified and fixed in advance. No variable selection was performed and the full set of predictors fit (see comments \cite{Marshall} on pre-specification of covariates in predictive modeling, page 2). There were 14 predictors in total for the CRT data and 8 for the CLL data.

To simplify the methodological and data-analytic development, we restrict ourselves in this paper to early death within a fixed time-window following patient study inclusion. This allows us to simplify to the analysis of binary outcome and logistic regression. Censored observations are treated as non-events.  For the CRT data,  we consider the first two years of follow-up,  for which we have 153 deaths and 38 censored records (3.6\%). For the CLL data,  we only investigate one-year survival where we have 184 early deaths and 46 censored records (6.6\%).

\section{Application and results}
We applied approaches 1, 2 and 3 to both the CRT and CLL data. Each approach was applied using either $K=1$ (single imputation), 10, 100 and 1000 as number of imputations. In addition, to allow for an assessment of  variation due to imputation, we repeated each application by generating 10 replicate analyses for each choice of $K$. We consistently used $L=10$ (number of cross-validation folds) throughout.
Within any application of a method, we calculated the final predicted probability of the binary outcome using both the mean and the median across the $K$ calibrated probabilities within an individual (note the latter will be constant by definition for completely observed reords in approaches 2 and 3). As we found very little difference between either the mean or median-based results,  we decided to only present mean-based summaries in this paper. To pool regression coefficients in approaches 2 and 3, Rubin's rule (mean averaging) was used.
Note that all approaches coincide for $K=1$.

All analyses were carried out using {\tt R} (3.4.3) \cite{R}.  Multiple imputations were generated using the package {\tt MICE} (2.46) \cite{mice} using chained equations and standard settings.   \cite{vanBuuren2015}

\subsection{Summary measures}
We focus on accuracy as measured by the Brier score (see \cite{Hand1997}, section 6.5,  page 107)  as well as a variance measure which is introduced below,
to compare performance between approaches on the real data.

The Brier score is calculated for each $r^{th}$ replication of an analysis with any approach for a fixed choice of $K$ as
\begin{equation}
B_r = \frac{1}{n}\sum\limits_{i=1}^{n} (\widehat{P}_{ir}-y_i)^2,
\end{equation}
with $\widehat{P}_{ir}$ the estimated event probability for the $i^{th}$ individual in the $r^{th}$ replicated analysis and $y_i$ the true class indicator. We average the 10 within-replicate Brier scores $B_r$, $r=1,...,10$ to obtain an estimate of the expected accuracy for the investigated approach at the number of imputations $K$.

The second summary is a measure of the amount of variation between the replicate predictions $\widehat{P}_{ir}$ for an approach with a fixed number of imputations $K$ and is defined as follows.
 We first calculate the mean prediction $\overline{P}_i$ across replications for each patient as well as the deviations
  $D_{ir}=\widehat{P}_{ir}-\overline{P}_i$.
While these deviations $D_{ir}$ are heteroscedastic,  their variation will be {\it approximately} constant across patients with $0.2\le \overline{P}_i \le 0.8$. We therefor discard all deviations corresponding to patients with $\overline{P}_i<0.2$ or $\overline{P}_i>0.8$ and compute the $90^{th}$ and $10^{th}$ percentiles $Q_{0.9}$ and $Q_{0.10}$ across all remaining deviations $D_{ir}$. Finally,  we report $R=(Q_{0.9}-Q_{0.10})*100$ as a measure of spread of predictive probabilities (expressed as percentage) induced by imputation variation at the probability scale.
While this variance measure is {\it ad hoc},  it has the advantage of providing an absolute measure of the change in predicted probabilities directly at the probability scale.  It is not affected by choice of transformation,  such as variance stabilizing transform and the need to back-transform to the original scale.

We calculated the above measures for both datasets and for $K=1$ (single imputation), 10, 100 and 1000. For the Brier score, the calculation was carried out on the full dataset, in addition to a calculation using only samples containing missing values and likewise using the complete observations only.
For the variance measure $R$, we calculated the measure separately on the complete cases,  and for observations containing missing values.

\subsection{Accuracy results}
\begin{figure}[!htbh]
\centering
\includegraphics[height=7cm]{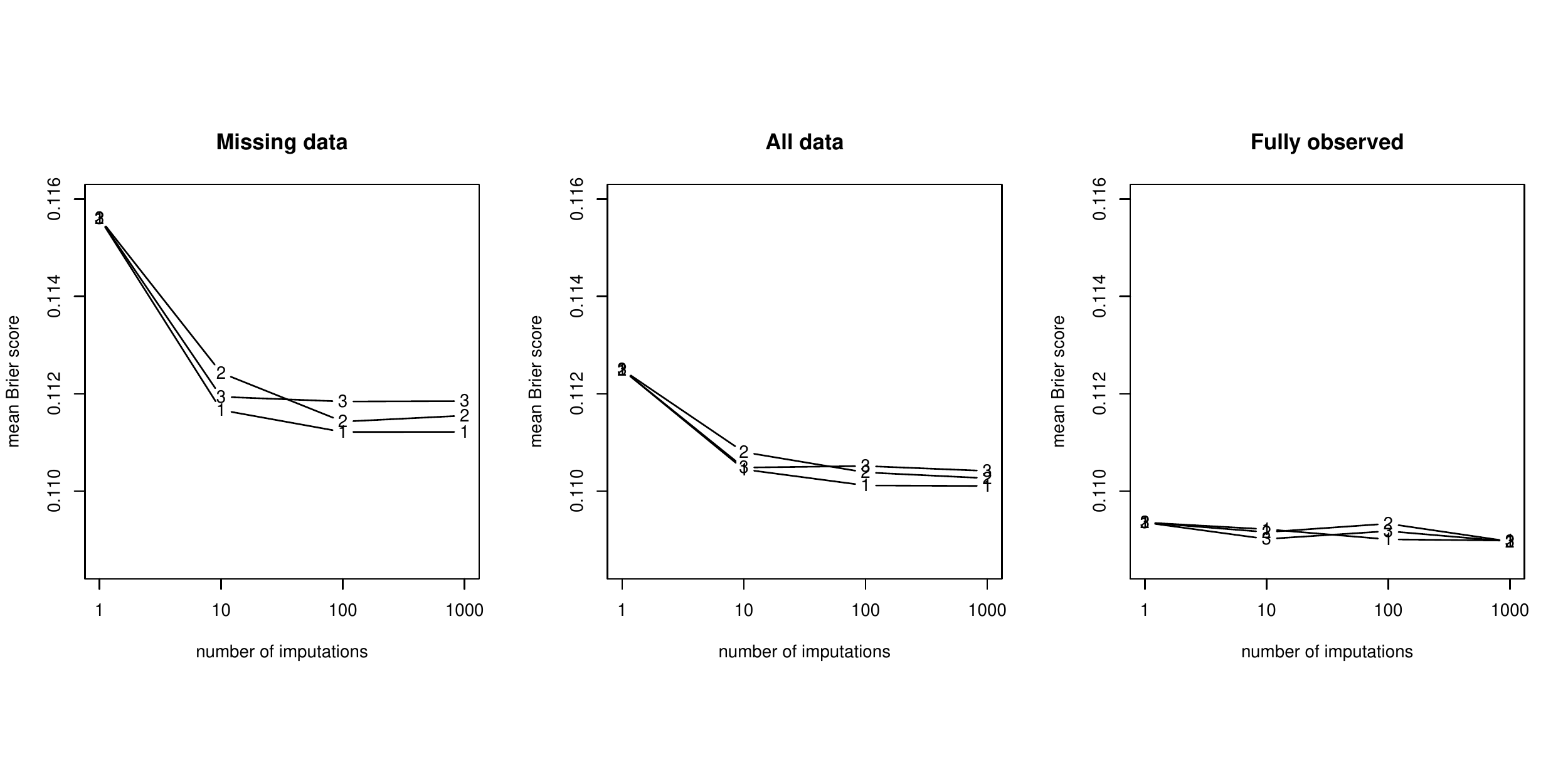}
\caption{\label{fig1} Average Brier scores for approaches 1, 2 and 3 calculated on the CRT data example, plotted versus the number of imputations used ($K$=1,10,100,1000). Results are presented as calculated on the full data set (middle plot),  using records containing missing values only (left-side plot) and using the complete cases only (right-side plot). The plotting symbol (1,2 or 3) indicates the approach used.
}
\end{figure}

Figures~\ref{fig1} and \ref{fig3} display results for Brier scores. The different plotting symbols 1, 2 and 3 distinguish between the 3 approaches.
As expected,  Brier scores are always higher when calculated on records containing missing values,  due to the greater uncertainty induced through the need to estimate these in imputation. Results calculated from the complete data are a compromise between Brier scores on the fully observed cases and those for records containing missing data.

Crucially,  for accuracy,  results do not seem to differ between the approaches,  whether investigating the CRT or CLL data. For the CRT data,  we notice a small decrease in Brier scores from $K=1$ to $K=10$ in both the missing data and when calculated across the entire dataset.  The same effect cannot be seen in the fully observed part of the data,  which indicates that the slight gain in accuracy is due to the increased precision gained by multiple as opposed to single imputation.  There does not seem to be further gain when increasing imputations beyond $K=10$ however.
In comparison and for the CLL data, Brier scores are essentially constant across $K$.

\subsection{Variation results}
\begin{figure}[!htbh]
\centering
\includegraphics[height=7cm]{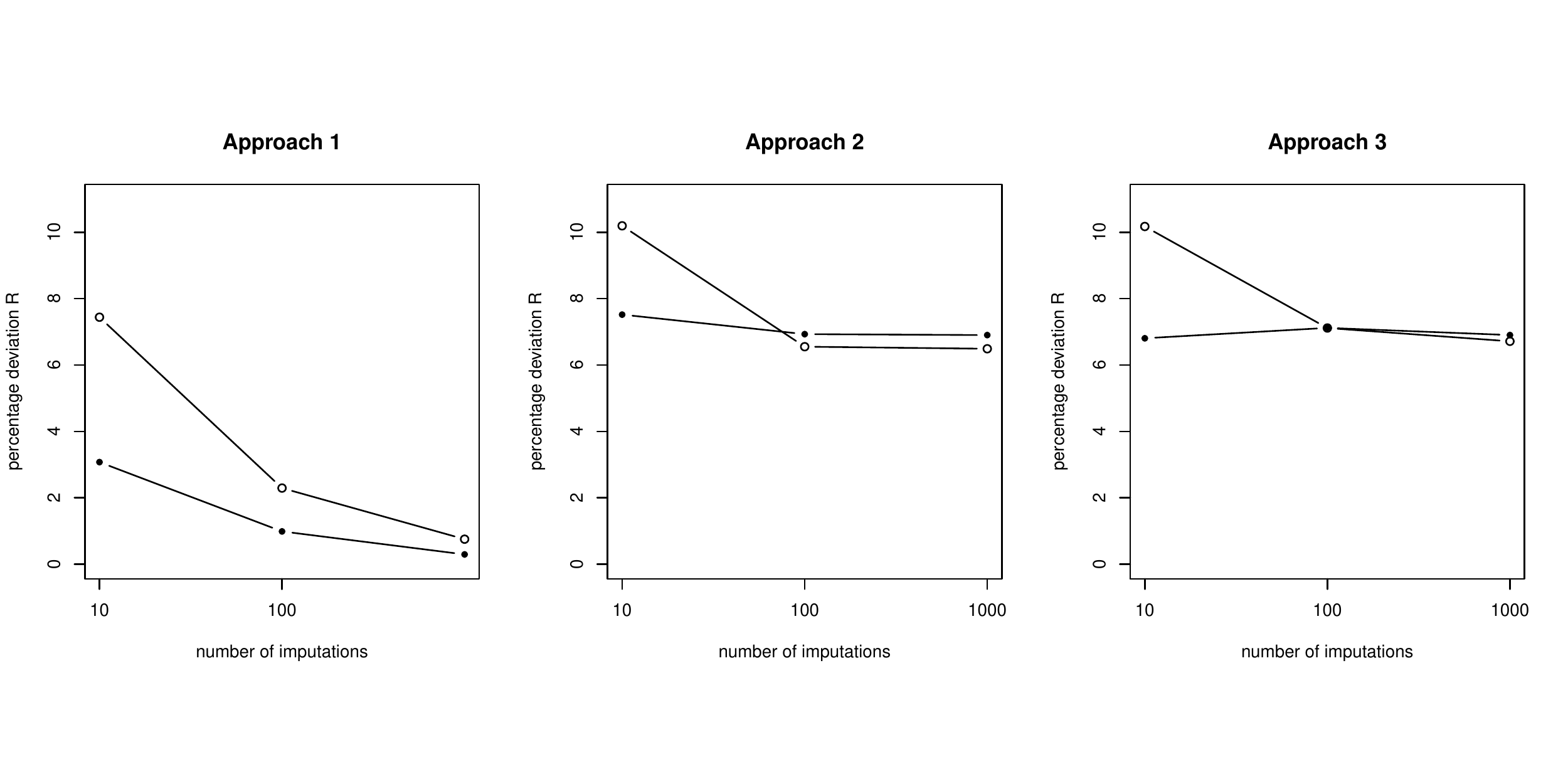}
\caption{\label{fig2} Percentage deviations of predictions $R$ across replicate calibrations for approaches 1, 2 and 3 in the CRT data example, plotted versus the number of imputations used ($K$=10,100,1000). Results are shown separately for fully observed records (solid dots) and observations containing missing observations (open dots). R measures at $K=1$ are 9.9\% for fully observed records and 20.6\% for missing observations and identical across approaches (hence not shown in above plots).
}
\end{figure}

\begin{figure}[!htbh]
\centering
\includegraphics[height=7cm]{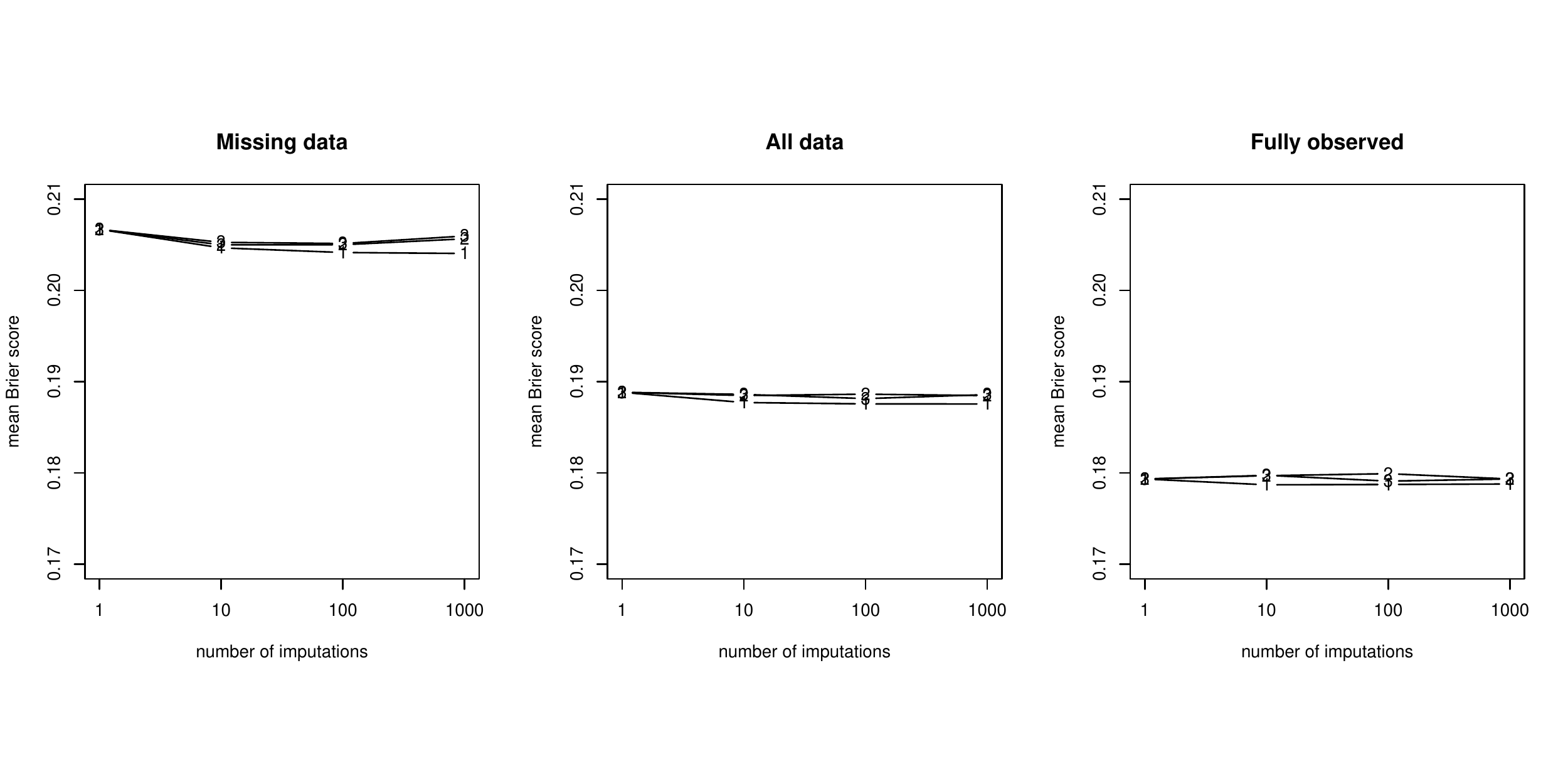}
\caption{\label{fig3} Average Brier scores for approaches 1, 2 and 3 calculated on the CLL data, plotted versus the number of imputations used ($K$=1,10,100,1000). Results are presented as calculated on the full data set (middle plot),  using records containing missing values only (left-side plot) and using the complete cases only (right-side plot).
The plotting symbol (1,2 or 3) indicates the approach used.
}
\end{figure}

\begin{figure}[!htbh]
\centering
\includegraphics[height=7cm]{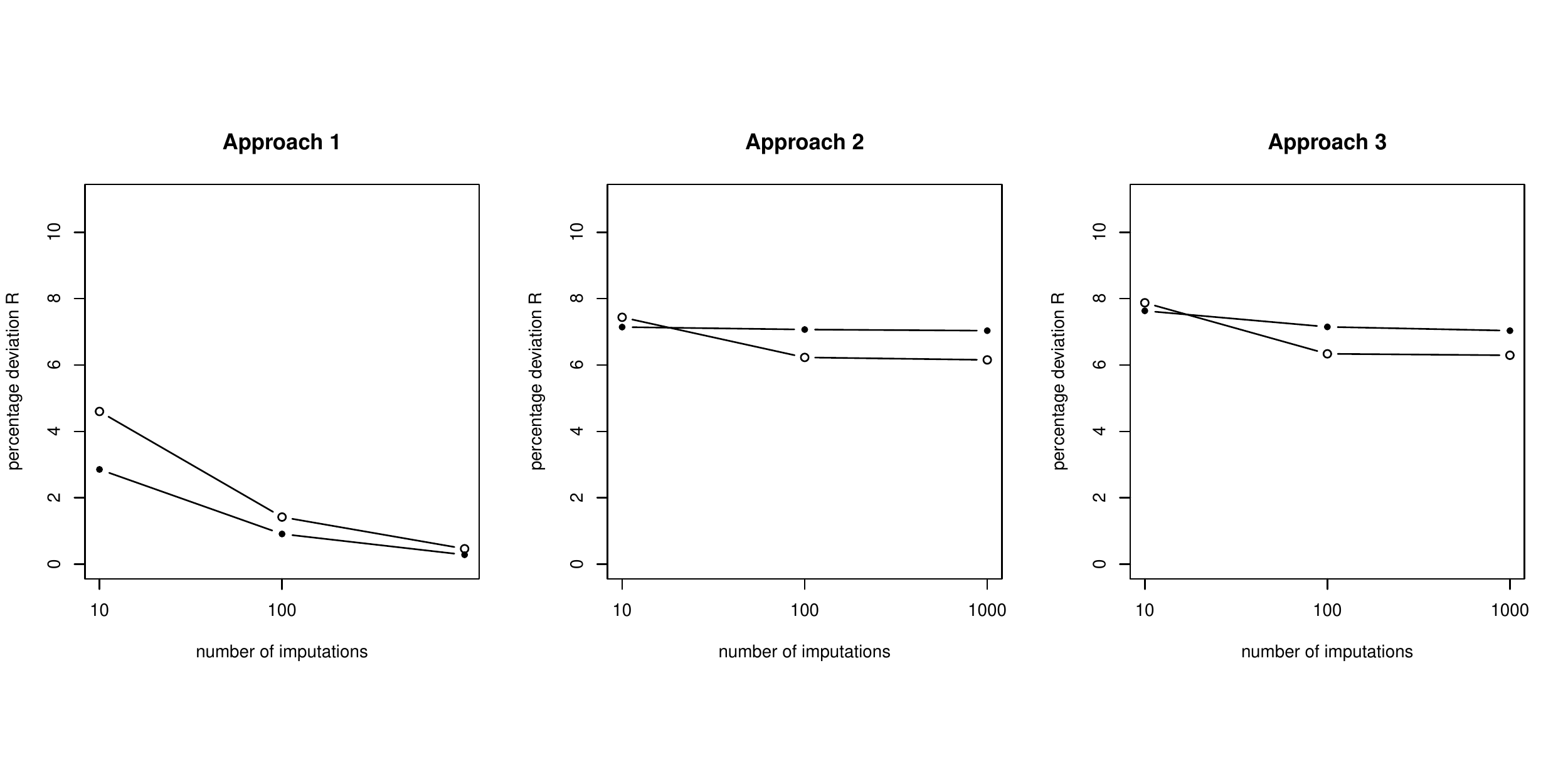}
\caption{\label{fig4} Percentage deviations of predictions $R$ across replicate calibrations for approaches 1, 2 and 3 in the CLL data example, plotted versus the number of imputations used ($K$=10,100,1000). Results are shown separately for fully observed records (solid dots) and observations containing missing observations (open dots). R measures at $K=1$ are 9.6\% for fully observed records and 15.3\% for missing observations and identical across approaches (hence not  shown in above plots).
}

\end{figure}

Calculating the variation measure $R$ for $K=1$, corresponding to single imputation and for which all 3 approaches coincide, gives $R=20.6$\% and
$R=9.9$\% when predicting with either partially observed records or the fully observed data respectively in the CRT data.  For the CLL data these numbers are $R=15.3$\% and $R=9.6$\% for partially and fully observed records respectively. As expected, predicting from fully observed records is ``more easy" in the sense that it is associated with less variability,  which is a consistent feature of the full results for $K=10, 100,100$ shown in figures ~\ref{fig2} and \ref{fig4}. It is due to prediction for fully observed records not being affected by the variation induced by the need to estimate the unobserved predictors using imputation as for the partially observed records, in addition to the variation in regression coefficients induced by imputation. The most striking feature may however be the magnitude of the absolute deviations among predicted probabilities fitted for $K=1$ and which occurs between the $10^{th}$  and $90^{th}$ percentile, due to imputation variation alone.

Figures \ref{fig2} and \ref{fig4} show the change in the variation measure $R$ when increasing $K$. The behaviour is very different between approach 1 versus approaches 2 and 3. For the CRT data, increasing $K$ to 10 imputations leads to a reduction of the variation measure to 7.4\% and 3.1\% for partially and fully observed data respectively.   These numbers gradually further decrease as we increase $K$ to 100 and 1000. Specifically $R$ reduces from 7.4\%, to 2.3\% and 0.8\% for partially observed data.  The reduction is from 3.1\% to 0.9\% to 0.3\%.

We can note how the variation measures reduce similarly for approaches 2 and 3 with increasing $K$, but very differently from approach 1. First note how an increase to $K=10$ reduces $R$ to 10.1\% and 7.5\% only for approach 2. Further reductions as we increase to $K= 100$ and 1000 are much smaller,  as we have $R=$ 6.6\% and 6.5\% for partially observed records at 100 and 1000 imputations respectively.
Similarly we have $R=$ 6.9\% and 6.9\% for fully  observed records at 100 and 1000 imputations.
Results from approach 3 are virtually identical.

Only for approach 1 do we observe a gradual decrease in variation of calibrated predictions as $K$ increases and as one would reasonably expect.  For approaches 2 and 3 the gains are however much smaller and non-existent once we have reached $K=100$,  after which no further reductions in variation is observed.  For any given level of $K$, approach 1 beats approaches 2 and 3 in terms of variation and for both fully and partially observed data. Note how the variation measures at $K=1000$ for approaches 2 and 3 are barely improving on the variation we can observe at $K=10$ for approach 1 already.

For approach 1 we can note that the payoffs for increased imputation face diminishing returns, although the variation continues to reduce towards the zero lower bound.  It is of interest that only for $K=1000$ and approach 1, variation is reduced to levels which may be acceptable for clinical application.

Results from the analysis of the CLL data (figure \ref{fig4}) are from a qualitative point of view a complete confirmation of the above observations. At $K=10$ and approach 1 for example,  $R=$ 4.6\% and 2.9\% for partially and fully observed records, with further reductions for increasing $K$ more modest but with variation gradually approaching zero. For approaches 2 and 3,  these numbers are 7.4\% and 7.1\% (approach 2)  and 7.8\% and 7.6\% (approach 3) and with negligible further reductions as $K$ is increased to 1000.  In fact,  for the CLL data,  variation measures $R$ are completely separated between approach 1 on the one hand and approaches 2 and 3 on the other.  The lowest variation measure $R=6.2$ at $K=1000$  for approach 2,  substantially above $R=4.6$\% for partially observed data for approach 1 with $K=10$.
Again we only achieve variation $R$ levels of 0.5\% and 0.3\% at $K=1000$ for approach 1, which again indicates that imputation numbers may need to be substantially increased beyond current practice.

Finally concerning approach 3, we note that neither gain nor loss of performance is observed relative to approach 2 in terms of accuracy and variance.  Importantly however, this also implies that using the mean imputation in prediction does not reduce the performance deficit relative to approach 1.

\subsection{Simulation}
We have carried out a simulation experiment to confirm some of the findings observed in the above data analytic application. The simulations are inspired from the variance-covariance structures observed in the CRT data. Both MAR and MCAR scenarios are investigated. A key advantage offered by the simulation is that it also allows us to explicitly investigate bias,  in addition to the variation measures we investigated above.

In brief, these simulations confirm and further support our above findings.  Specifically, we could not find any evidence of bias in predicted probabilities with any of the investigated approaches.  Results however clearly confirm much lower variation measures from approach 1 as compared to approach 2, as we found in the data application.
We will make these materials and code available online as supplementary material to the paper.

\section{Conclusions and discussion}
We have investigated the problem of combining predictive calibration with (cross) validation in prognostic applications, when multiple imputations are used to account for missing values in predictor data. Instead of following a primarily algorithmic ad hoc approach, we have commenced with a review  of the theoretical foundations of multiple imputation in the predictive setting. Specifically, we clarified how predictive calibration requires estimation of a different predictive density (equation~\ref{eq3}) - and thus integration across both missing observations and unknown effect parameters - as opposed to averaging across missing values only (equation~\ref{eq1}) which is implicitly implemented in current standard multiple imputation software.
Instead of pursuing a direct, fully Bayesian approach to the calculation of the integrals as in \cite{Erler2015} and \cite{Erler2017}, we have proposed a methodology which estimates by approximation the expectation of the required predictive density. We achieve this by averaging the predictions from individual models fitted on the single imputed datasets within a set of (multiple) imputations that can be generated with existing multiple imputation software
(approach 1). We contrast this methodology with direct use of Rubin's rules-based model calibrations (approaches 2 and 3).
Finally, we compared methods on accuracy and variance measures calculated on cross-validated estimates of the predicted probabilities in two real data sets, as opposed to simulations.

Results suggest that methodological approaches are indistinguishable with respect to accuracy (root mean squared error). We suspect this result may well extend to bias as suggested by a limited simulation exercise. Large differences from both the qualitative and quantitative point of view are however observed between approach 1 (combining predictions) and approaches 2 and 3 (pooling regression coefficients) with respect to the variation of  predictions for individual patients between repetitions of the procedure with different imputations (analysis replication). The following observations can be made.
\begin{enumerate}
\item Absolute levels of variation of predicted probabilities are very high when using single imputation.
\item Multiple imputations must be used to reduce this variation, but approach 1 is vastly more efficient in variance reduction as compared to approaches 2 and 3 for the same increase in imputation numbers.
\item Only approach 1 appears to have the basic property of variation approaching zero as the number of imputations increases. For approaches 2 and 3, variance measures stabilize once 100 imputations have been used and do not reduce further.
\item Numbers of imputations used in predictive modelling may need to be drastically augmented above current  clinical practice to reduce variation to levels suitable for routine clinical application.  Numbers closer to 1000 or beyond imputations may be required. A literature review (\cite{Marshall}, page 6) indicates the majority of clinical applications used between 5 and 10 imputations.
\item Use of single imputation in predictive calibration should be
rejected and the practice phased out.
\end{enumerate}

Irrespective of the above results, we hope our paper would stimulate the medical statistical community to propose future work on combination of multiple imputation, predictive calibration and validation by clear reference to the theoretical background  as we have tried to do in this paper - instead of the ad hoc algorithmic approaches which dominate the recent literature.
%Many of them seem to have no connection to prediction or prognostic context - or could in any sense lead to a reasonable or reliable assessment of the performance.
In principle, pursuing a fully Bayesian approach would ensure such rigour as it automatically leads to calibration of the required integrals described in section 2.
An alternative might be to adapt existing multiple imputation software such that it allows to save the imputation model equations for use in the imputation of future observations which may have missing predictor values.
Ideally such software would also incorporate modelling of the substantive outcome to be predicted,  such that both objectives can be achieved simultaneously.
Current multiple imputation software is, to our knowledge, focused on estimation of (pooled) regression (effect) measures and standard errors within a fixed dataset.

An additional objective we hope our paper would stimulate is to entice the medical statistical community to evaluate model approaches {\it \bf on real data} and data-based summary statistics such as accuracy or direct measures of variance as in this paper - as opposed to simulations which can too easily be subtly manipulated or selected to suit researchers needs or preconceived ideas.
Current literature on predictive calibration and validation with imputation typically reverts to simulations, sometimes presented as ``data-based" simulations.
In addition, we hope that greater attention will be placed on the assessment of predictive variation.  Much of current literature focuses too narrowly on assessment of bias. From the predictive point of view and when imputation is used,  variation may be at least as important if not more.

We conclude with a number of smaller remarks to point out connections with the wider literature and application field.
The first is the obvious connection between machine learning, particularly ensemble learning, and approach 1. It is known that ensemble methods (\cite{Breiman1996}; \cite{Wolpert1992}; \cite{Hastie2008}, section 8.8) can be highly effective at variance reduction through averaging of predictions from multiple constituent models.  The latter are usually obtained through re-fitting of some base-model, often after perturbation of the data in some sense,  such as bootstrapping.  In our case the perturbations can be thought of as arising form distinct realisations  of the required imputation.

We have consistently used means to pool the predicted probabilities (as for the regression coefficients using Rubin's rule) in this paper.  One could however imagine other choices,  such as averaging at the logit-scale, or use of medians and so on,  which would not substantially alter the nature of the approaches shown.
For the research in this paper we have recalculated all results using medians and found results which are both from a quantitative, and hence also qualitative, point of view near-identical to the results shown here. We also visually inspected the distributions of probabilities averaged and found them to be near-symmetric near the mode. To simplify the presentation we therefor decided to use the mean throughout.

Our paper has focused on logistic regression for binary outcome in prognostic studies. To achieve full generality however, the extension to life-time outcomes in the presence of censoring should also be investigated,  particularly for Cox models. This entails some special complications, apart from censoring, particularly the need to also address variation in baseline hazards as well as special considerations as to how censored survival outcomes should be accounted for within multiple imputation (\cite{Carpenter2013}, chapter 8).
We have carried out this research and can confirm that the key results from this paper on bias and variance  completely carry over to the survival setting with Cox regression analysis as well.
The description of this research however requires a separate dedicated paper.

\mbox{}
\\
\mbox{}
\\
\noindent {\bf Acknowledgements}\\
\mbox{}
\\
We thank Johannes Schetelig (University Hospital of the Technical University Dresden/DKMS Clinical Trials Unit, Dresden, Germany) for making available the CLL dataset and for the extensive joint work on the previous analyses. We thank Ulas Hoke and Nina Ajmone (Dep. Cardiovascular Research, LUMC, The Netherlands) for the opportunity and fruitful collaboration working on the CRT data. We also thank EBMT and DKMS for their work in collecting and preparing the CLL data. We acknowledge John Zavrakidis, Ali Abakar Saleh and Theodor Balan for their valuable contributions.

\newpage

\newpage
\bibliographystyle{Plain}

\newpage

\end{document}